\documentclass[manuscript]{acmart}
\usepackage{subcaption}

\AtBeginDocument{%
  }

\setcopyright{acmlicensed}
\copyrightyear{2025}
\acmYear{2025}
\acmDOI{XXXXXXX.XXXXXXX}
\acmISBN{978-1-4503-XXXX-X/2018/06}

\begin{document}

\title{The Evolution of Coordination in a Collective Intelligence System: 25 Years of English Wikipedia and the Emergence of Generative AI}


\author{Neal Reeves}
\email{neal.t.reeves@kcl.ac.uk}
\orcid{0000-0002-1044-3943}
\affiliation{%
  \institution{King's College London}
  \city{London}
  \country{United Kingdom}
}

\author{Maja Świeczkowska}
\email{maja.swieczkowska@kcl.ac.uk}
\orcid{0009-0001-6682-7057}
\affiliation{%
  \institution{King's College London}
  \city{London}
  \country{United Kingdom}
}

\author{Amy Rechkemmer}
\email{amy.rechkemmer@kcl.ac.uk}
\orcid{0000-0001-7572-751X}
\affiliation{%
  \institution{King's College London}
  \city{London}
  \country{United Kingdom}
}

\author{Elena Simperl}
\email{elena.simperl@kcl.ac.uk}
\orcid{0000-0003-1722-947X}
\affiliation{%
  \institution{King's College London}
  \city{London}
  \country{United Kingdom}
}

\renewcommand{\shortauthors}{}

\begin{abstract}
English Wikipedia is one of the largest examples of collective intelligence on the Web, sustained not only by article production but also by volunteer coordination and governance. While prior research has examined coordination work in Wikipedia, less attention has been paid to how participation in these spaces has evolved over time. Drawing on a longitudinal analysis spanning nearly 25 years of English Wikipedia, we examine editing patterns across five namespaces covering content, discussion, and governance. We find that participation in coordination spaces has declined relative to content production, particularly in governance areas, with a shrinking core of editors performing an increasing share of this work. Using Markov-based session metrics, we also find that editing has become more specialised, with editors moving less frequently between namespaces. Motivated by recent governance debates around generative AI, we conclude by investigating whether the availability of LLMs has altered these long-term trends. While short-term changes are visible, we find little evidence that generative AI fundamentally changed existing trajectories of coordination and participation.
\end{abstract}




\received{15 June 2026}

\maketitle

\section{Introduction}

Wikipedia is one of the largest examples of human collaborative effort on the Internet with over 65 million articles~\footnote{\url{https://en.wikipedia.org/wiki/Wikipedia:Size_of_Wikipedia}} in over 340 languages~\footnote{\url{https://meta.wikimedia.org/wiki/List_of_Wikipedias}} as of 2025. Founded in 2001 on the principle that ``anyone can edit'' content \cite{yarovoy2020assessing}, Wikipedia is maintained by a dedicated volunteer user base \cite{yarovoy2020assessing}. As well as serving as a crucial information gathering resource, prior research has highlighted the important role that Wikipedia content plays in other online peer-production communities \cite{vincent2018examining}. Moreover, Wikipedia's importance to web infrastructure and role as a ``gateway to the web'' have been well documented \cite{piccardi2021value}, and the platform is now seen as a case study of ``good faith collaboration'' in online spaces~\cite{reagle2010good}.

Wikipedia's collaborative efforts can be classified as either direct or indirect coordination, with both forms contributing to the success of the platform. This coordination takes place across various environments for engagement, known as \textit{namespaces}, that serve different purposes and allow for coordination in different ways. Within the main article editing namespace, for example, indirect coordination can be found in the form of stigmergy ~\cite{arazy2020emergent}. This is shown through increased engagement as editors are driven to directly contribute to the content -- or digital traces -- provided by others~\cite{zheng2023stigmergy}. Meanwhile in other namespaces, 
%often unbeknownst to casual readers~\cite{ren2023did}, 
direct coordination takes place through talk pages of editor discussions regarding articles and governance of Wikipedia as a whole. These discussions often represent editing suggestions~\cite{bipat2018we}, playing a crucial role in guiding and shaping article content. As such, these are often classified as a form of articulation work\footnote{The processes and tasks needed to facilitate work, which may themselves consist of work \cite{lee2015matrix}} on the platform. Though not directly contributing to article content, articulation work on Wikipedia is considered a valuable, yet under-appreciated service by editors~\cite{kriplean2008articulations}, and it has been found to contribute positively to article quality~\cite{crowston2020effects}.

%%The bulk of Wikipedia content lies in articles which are maintained through a form of stigmergic coordination where editors collaborate directly through the process of editing articles \cite{arazy2020emergent}. This form of collaboration has been linked to increased levels of engagement as participants are driven to work on editors after seeing the content -- or digital traces of -- other participants' contributions to edits \cite{zheng2023stigmergy}. Moreover, stigmergic coordination of editing has been associated with the greatest increases in article quality \cite{crowston2020effects}. Nevertheless, other engagement environments in Wikipedia known as \textit{namespaces} also play a crucial role in guiding and shaping content. For example, in English Wikipedia, a plurality of article talk page content is used to coordinate and make editing suggestions \cite{bipat2018we}. While namespaces are intended to support and guide collaboration between users, most users read only the main Wikipedia namespace \cite{ren2023did}.

Despite the importance of this coordination layer, we know relatively little about how it has changed as English Wikipedia has matured. Recent research has highlighted the lack of longitudinal analyses exploring the evolution of Wikipedia's coordination activities \cite{ren2023did}. While Wikipedia's declining user base is well documented, what this means for articulation and coordination work -- as well as how the editing community has responded to it -- is less clear~\cite{smith2020keeping}. A further question is how this coordination work has evolved in response to changes in the platform. Among the developments in Wikipedia in recent years, the emergence of Large Language Models (LLMs) is among the most consequential, contributing to the transformation of content and user engagement~\cite{lyu2025wikipedia,reeves2024exploring}. Concerns surrounding editor engagement on the platform have prompted discussions about the ways in which LLM usage could impact the relationships within the closely collaborative environment, highlighting that writing bots may decrease contributions~\cite{wagner2025death}, which could have downstream effects across namespaces. Editors have engaged in a long and at times heated series of discussions around the use of LLM content in Wikipedia, culminating with the decision to block such content from English Wikipedia in March 2026\footnote{See \url{https://en.wikipedia.org/wiki/Wikipedia:Artificial_intelligence\#Discussion_timeline} for a full timeline.}.

To that end, we present -- to our knowledge -- the first longitudinal evaluation of Wikipedia engagement across five namespaces covering main article editing, coordinated collaboration and governance. We characterise these namespaces as \textit{articulation spaces} \cite{boden2014articulation} where direct and indirect coordination of work takes place.

We frame our evaluation around the following research questions:
\begin{enumerate}
    \item How has the scale of coordination changed relative to content editing over Wikipedia’s history?
    \item How do session-level editing behaviours reveal specialisation and transitions across collaboration spaces?
    \item How has the emergence of LLMs reshaped engagement in articulation spaces compared to content spaces?
\end{enumerate}

By considering article editing patterns as \textit{edit sessions}, we explore how the relationship between engagement in different namespaces has evolved, as well as how editors have transitioned within and between namespaces over almost 25 years of Wikipedia. Additionally, by using an Interrupted Time Series Difference-in-Differences approach, we explore how engagement in coordination namespaces has changed relative to engagement in the main article namespace following the release of ChatGPT on October the 31st 2022.

\section{Related Work}

%We are far from the first to analyse Wikipedia engagement. 

In this section, we outline prior research across three key topics: coordination in Wikipedia, Wikipedia engagement, and the impact of LLMs on Wikipedia.  

\subsection{Articulation Work in Online Spaces}
%Despite its importance for online communities, such work is often overlooked or underappreciated, causing the contributors who fill such roles to leave communities due to stress and burn-out \cite{linaaker2024sustaining}.

Articulation work describes the meta-activities that facilitate the functionality of socio-technical systems \cite{huber2023navigating}. This work is typically characterised as ``invisible'' labour \cite{gruszka2022out} that is necessary for moderating and managing content \cite{meluso2024invisible}. Findings from virtual Citizen Science communities have highlighted the important role that recognition can play in motivating continued participation in both articulation and core work \cite{boone2024reimagining}.

%Findings from open-source software communities have highlighted how the most active participants increasingly take on major coordination and community management roles \cite{geiger2021labor}.

%A large body of research involving articulation work has focused on online community moderation. \citet{cai2022coordination} analysed how volunteer moderators in the Twitch platform collaborate to coordinate the articulation work necessary to manage and facilitate live streaming. \citet{jiang2019moderation} have highlighted how the emergence of novel affordances (such as voice chat) have led to challenges and increasing workloads for moderators and those who engage in moderation tasks. Recent work by \citet{almeda2025creativity} has similarly surfaced how community support activities in creative online communities are being disrupted -- and perceived -- by generative AI technologies. Our work partially responds to these topics in more depth by analysing the impact of generative AI on Wikipedia coordination and articulation work. Nevertheless, we caution that Wikipedia differs from these other online communities as ``anyone can edit,'' and so there is less of a focus on traditional moderator-moderated structures. 

We note work which has focused on Wikipedia moderation and articulation tasks. \citet{tran2022risks} explored pre-publication moderation of Wikipedia articles across 17 language editions and found only a moderate impact on engagement. \citet{houtti2022we} explored discussions and methods for prioritising and evaluating Wikipedia articles through namespace discussions. These methods contribute to directing and coordinating editor effort, but fall prey to tensions between individual and collective priorities. \citet{im2018deliberation} analysed 7,316 Wikipedia Requests for Comments (RfCs) as a key form of governance decision-making. Their findings suggest that many RfCs are less effective than they could be due to editors lacking confidence in closing and acting on these requests.\par
While these works all provide insight into coordination work within Wikipedia, they each focus on specific subtypes of coordination. We build on this existing body of work by taking a broader, longitudinal lens and exploring the relationship between coordination namespaces and overall edit activity over time.

\subsection{Coordination on Wikipedia}
Wikipedia's role as a dispersed knowledge repository has prompted efforts to study the ways in which users coordinate work. Stigmergic collaboration has been examined using clustering to visualise where editors are 'excited' by each other's contributions and return to build on them in the future \cite{zheng2023stigmergy}. Both \citet{zheng2023stigmergy} and \citet{crowston2020effects} demonstrate the improvement of article quality in environments where stigmergic collaboration was high. This form of collaboration has also been seen  to provoke opinion clashes, leading to edit wars, where editors overwrite other user's edits to submit their version of a change to an article \cite{chhabra2020dynamics}. The dynamic of these edit wars has been researched by looking at revision histories of notable controversial articles and their number of edits, reverts per revision \cite{halfaker2011don}, and amount of content removed \cite{chhabra2020dynamics}.\par
Although direct collaboration through \textit{talk} pages linked to articles has shown a smaller increase in quality compared to stigmergic collaboration \cite{crowston2020effects}, clustering articles based on quality and using metrics measuring explicit and implicit collaboration \cite{zhang2020mining} revealed that higher quality article pages often had active talk pages, noting the importance of explicit collaboration further down the life cycle of an article. A mixed methods approach to studying the Wikiprojects pages pointed out that projects with higher relationships, measured by the amount of messages, often signified higher quality end products within those projects \cite{rychwalska2021communication}. Network analysis of contributors on both low and high quality articles also revealed high quality articles contain more nodes with more than one connection to other author nodes \cite{de2015measuring}. Although the form of collaboration is not explicit, the relationship between close collaboration and article quality is clear.

\subsection{Wikipedia Editor Engagement}

%%A wide body of research has explored engagement in Wikipedia across a number of metrics and media. \citet{piccardi2023large} explored how users browse Wikipedia through an evaluation of four weeks of English language server logs. Their findings suggest engagement is closely linked to overall article quality and that readers abandon reading when faced with low quality content. Evaluations of engagement with Wikipedia content have focused on diverse article content such as images \cite{rama2022large}, references \cite{piccardi2020quantifying} and links to external content \cite{piccardi2021value}. 

\citet{lanamaki2018latent} performed a longitudinal analysis of group formation and behaviours among Wikipedia editors, while \citet{das2022quality} performed a longitudinal evaluation of quality change in Wikipedia articles finding that 50\% of all articles experience no long-term improvement in quality. Similar to our focus on Wikipedia change over time, \citet{matei2017wikipedia} analysed how Wikipedia had evolved, identifying key trends and distinct stages in this evolution. However, we note that each of these sources focuses largely on the main article namespace. Our work provides further insight into Wikipedia's evolution by surfacing the relationship between article editing and coordination work as well as exploring how use of these namespaces has changed over time.

%%\subsection{Categorising Editors}
%%The behaviour of editors has been categorised using various techniques into six clusters \cite{liu2011does}, and more recently, eight distinct roles \cite{yang2016did}, based on the types of edits and interactions they perform. In both cases it was generally agreed that higher quality articles will contain many varying contributions from a wide range of roles, although some roles became associated with a decrease in quality at different points in an article's lifecycle \cite{yang2016did}. A study done on the Chinese edition of Wikipedia identified that altruism and social belonging strongly motivated community contributions, while content contribution was motivated by self-development and enjoyment \cite{xu2015empirical}. Further research has revealed that these different motivations, when applied to new editors, can predict patterns of contributions across different Wikipedia namespaces \cite{balestra2016motivational}. With concerns to life span, evidence suggests new editors may be more inclined to leave the community based on their contributions being reverted \cite{halfaker2011don}, of which there is an increase after spikes in attention towards Wikipedia \cite{zhang2019participation}. This is in addition to experienced editors departing due to interpersonal conflicts causing burnout \cite{konieczny2018volunteer} or bullying from other editors \cite{das2021expertise}, as identified through user's talk pages. 

\subsection{The Impact of LLMs on Wikipedia}

Page views analysed through Difference-in-Differences methods \cite{lyu2025wikipedia, singh2024impact}, loosely suggest a decrease for articles after the release of ChatGPT. Pages classified as having similar outputs to ChatGPT often experienced a decrease, compared to pages with dissimilar outputs \cite{lyu2025wikipedia}. Both studies suggested further investigation to solidify the results, giving the opportunity for analysis of these metrics utilising different methodologies. In addition, Huang et al. \cite{huang2025wikipedia} analysed page views across different topics on Wikipedia such as \textit{Art} and \textit{Computer Science}, revealing minor recent drops in views across some scientific topics, however, noting that the direct impact of LLMs is still uncertain. A feedback model was suggested that links viewing and contribution to Wikipedia's asset value and search trends \cite{wagner2025death}, identifying the increase in non-human viewers and editors, predicting their further increase post the mainstream use of LLMs for content generation. \par
Interviews with Wikipedia editors and domain experts involved with Wikimedia, have suggested that content generation tools may improve productivity, but also proliferate misinformation \cite{fordimplications} and threaten Wikipedia's collaborative environment \cite{vetter2025endangered}. Such mixed results are also reflected in community participation~\cite{zhou2025}, where LLMs lower entry barriers for new editors while at the same time raising the standards of contributions. These outcomes prompt further research into the influence of LLMs on Wikipedia's collaborative ecosystem, to understand how editor behaviour has changed since, and the evolving balance of work on the platform.

\section{Data and Methods}

\subsection{Data}

To perform our analysis, we used Wikipedia dump files consisting of a total of 18,241,443,327 edits made across the entire lifespan of the English Wikipedia starting on the 21st of January 2001 and ending on the 21st of July 2025. 

\subsection{User Processing and Bots}

As well as human users, Wikipedia has relied heavily on the engagement of bots which support editors by completing tasks ranging from fixing and protecting articles to generating article content \cite{zheng2019roles}. Bots make approximately 15\% of all Wikipedia contributions although a small number of bots make disproportionately large numbers of edits \cite{ren2023did}. Their inclusion has been noted to dramatically distort analyses of human edit behaviours \cite{hall2018bot}.

To identify bots we used four sources: 
\begin{enumerate}
    \item The user-groups dump made available by the Wikimedia Foundation; 
    \item The former user-groups dump; 
    \item A list of unflagged bots maintained by Wikipedia users\footnote{See \url{https://en.wikipedia.org/wiki/Wikipedia:List_of_bots_by_number_of_edits/Unflagged_bots}}; 
    \item The Wikipedia category ``All Wikipedia Bots''\footnote{See \url{https://en.wikipedia.org/wiki/Category:All_Wikipedia_bots}}. 
\end{enumerate}

We used the Wikipedia SQL instance\footnote{\url{https://quarry.wmcloud.org}} to convert usernames to user IDs. We recognise this list is unlikely to be exhaustive and we also noted discrepancies in resources. For example, while 2,164 bots are listed on the Wikipedia page for the ``All Wikipedia Bots'' category, only 1,710 user IDs could be retrieved from the system, perhaps because some had been exhausted or because accounts had not visited English Wikipedia.

%It should be noted that this list is unlikely to be exhaustive.  Global accounts will not appear on Wikipedia editions a user has not visited and it is therefore likely that many of these listed accounts have not edited the English Wikipedia. Even so, we caution that our method may not fully capture all bots, particularly those which have not yet been flagged or identified as a potential bot by the Wikipedia community.  

%To identify edit sessions, we also needed to be able to identify concurrent submissions from the same Wikipedia user. This necessitated that any submission without a user identifier be removed. While some otherwise anonymous users could be detected based on their IP address, 233,029,000 submissions were removed due to the lack of a user ID representing 1.28\% of all edits. 

\subsection{Session Definition}

To identify edit sessions, we grouped edits in chronological order for each user. We grouped edits into sessions with a new session determined to start when the period between edits exceeded 1 hour in line with prior work by \citet{geiger2013using}. Edit sessions are reported based on the date and time of the \textit{first} edit in a session in case of sessions spanning boundaries between months or years. To identify edit sessions, we required indications of Wikipedia users and therefore removed a total of 233,029,000 anonymous submissions representing 1.28\% of all edits.

%To identify edit sessions, we first grouped edits in chronological order for each user. We then divided these edits into groups based on the period between them where a new session was defined as starting once the time between edit n and edit n+1 for a given user exceeded a specific threshold. Following prior work by \cite{geiger2013using}, we used a cut off of 1 hour. While this likely exceeds the size of most sessions, we wished to ensure our approach was robust and would adequately reflect the behaviour of the small proportion of users who display the heaviest of editing behaviours. We define the time at which an edit takes place based on the \textit{first} edit within a session. Edit sessions which span month or year boundaries are thereby recorded only once to avoid duplication or splitting of sessions. 

\subsection{Namespace Selection}

Wikipedia features many diverse namespaces\footnote{See \url{https://en.wikipedia.org/wiki/Wikipedia:Namespace}}. For this analysis, we focus on the relationship between main article edits (namespace 0) and the most commonly used coordination and governance spaces as summarised in Table~\ref{tab:namespaces}. For brevity and clarity, we refer to ns 4 as governance and ns 5 as governance discussion as these largely reflect the main purpose of these spaces.

\begin{table}[ht]
\caption{Namespaces included in our analysis.}
\label{tab:namespaces}
\begin{tabular}{llll}
\toprule
Namespace & ID & Role in analysis & Purpose \\
\midrule
Main & 0 & Content & Encyclopaedic articles \\
Talk & 1 & Articulation & Discussion of individual articles \\
User talk & 3 & Articulation & Communication between editors \\
Project & 4 & Governance & Policies, guidelines, Requests for Comment, WikiProjects \\
Project talk & 5 & Governance discussion & Discussion of policies and project pages \\
\bottomrule
\end{tabular}
\end{table}

We exclude namespace 2 (user pages) as these pages include a sandbox functionality intended to allow editors to learn how to update Wikipedia. Unlike other namespaces, many edits are therefore not intended to be public-facing. Remaining namespaces primarily support technical infrastructure or functionality rather than coordinating editing work and were therefore deemed to be outside the scope of this research.

%Wikipedia features a diverse range of namespaces, although some namespaces are niche or have been retired\footnote{For the full list, see: \url{https://en.wikipedia.org/wiki/Wikipedia:Namespace}}. To explore the relationship between coordination and main article edits, we focus on the main article namespace 0 and four namespaces the intended for coordination work: talk (ns 1); user talk (ns 3); project (ns 4) and project discussion (ns 5). The project namespace is home to content describing the Wikipedia project itself and associated initiatives such as WikiProjects. For ease of readability, we refer to ns 4 as governance and ns 5 as governance discussion as these are largely the purpose of these namespaces. 

%We chose not to include namespace 2 (user pages) due to their function as experimentation and sandbox stages. Unlike other namespaces where edits are intended to be public, user profiles feature a sandbox where editors can safely learn how to edit Wikipedia \cite{kenny2013collaborative}. While we believe this is an interesting area of study in its own right, we note clear differences between private editing practice and public coordination activity and therefore excluded this namespace from our analysis. 

\subsection{Analysis}

\subsubsection{Gini Coefficient} The Gini Coefficient is a measure of resource inequality that can be understood as the percentage inequality in the share of a resource (in our case, edits) in a population (in our case, Wikipedia editors) \cite{catalano2009measuring}. The higher the coefficient value, the greater the inequality in the system such that 0 represents perfect equality and 1 represents perfect inequality. We calculate this metric using the following formula:

\[
G \;=\; \frac{\sum_{i=1}^{n}\sum_{j=1}^{n} \lvert x_i - x_j \rvert}{2n^2 \,\bar{x}}
\]

Where $x_i$ represents the observed values (i.e., edits per user), $\bar{x}$ is the mean and $n$ is the number of observations (i.e., number of users). 

\subsubsection{Session Level Markov Metrics} 

The questions we pose in RQ2 are properties of \textit{transitions} as opposed to edit volumes. Factors such as whether editors move freely between content, discussion and governance activities or remain consigned to a single namespace are largely invisible to edit counts. Instead, they require consideration of the extent to which and order in which namespaces are visited. To capture this, we operationalise edit sessions as a sequence of namespace states and analyse the resulting transition structure. In contrast to prior work using focused at the whole session level (e.g., \cite{geiger2013using}), we focus on analysing the within-session \textit{sequence} of transitions.

We operationalise edit sessions as Markov chains\footnote{A stochastic process consisting of a finite set of states and transition probabilities, where it is assumed the conditional distribution of the next state depends only on the present state \cite{ross2014}.} where each edit in a session corresponds to a state in the chain. From these chains, we calculate three metrics: stationary-weighted self-transition probability, spectral gap and mixing time.

%These three metrics capture transition at complementary scales. Stationary-weighted self-transition probability focuses on the likelihood of a session to remain within its existing namespace and is therefore focused on local effects for each namespace. Conversely, the spectral gap and mixing time are \textit{global} measures of how quickly the session as a whole ``forgets" the namespace in which it started. Combined, these separate local persistence from integration at the system level.

For each metric, $\{X_t\}_{t \geq 0}$ represents a finite Markov chain, where the state space is represented by $\mathcal{S}$ and the transition matrix is $P = (p_{ij})$, where $p_{ij} = \Pr(X_{t+1}=j \mid X_t=i)$. $\pi$ denotes the stationary distribution of the chain.

%\paragraph{Entropy Rate} This metric reflects the average uncertainty at each transition within the chain. A higher entropy represents a higher degree of randomness in the transitions and a lower degree of predictability for the next transition. In our analysis, this represents how predictable the editors next choice of namespace is based on their current choice of namespace. We calculate entropy using:

%\[
%H(P) \;=\; - \sum_{i \in \mathcal{S}} \pi_i \sum_{j \in \mathcal{S}} p_{ij} \, \log_2 p_{ij}.
%\]

\paragraph{Stationary-Weighted Self-Transition Probability}. This metric describes the probability that the chain will remain in the same state upon transition (i.e., that a user will continue to edit within the same namespace). For conciseness, we describe this metric using the term ``stickiness'' to represent how a user may get `stuck' within a given namespace during edit sessions. We calculate this probability using the formula: 

\[
Probability \;=\; \sum_{i \in \mathcal{S}} \pi_i \, p_{ii}.
\]

\paragraph{Spectral Gap} The spectral gap is a measure of diffusion within a system and represents the time taken to converge to the stationary distribution. In the case of our analysis, this represents the time taken for a chain of edits to converge with the longer-term distribution of edits across namespaces. The larger the spectral gap, the faster the convergence towards the stationary distribution. We calculate the spectral gap using:

\[
\gamma \;=\; 1 - |\lambda_2|
\]

Where $\lambda_2$ represents the second largest eigenvalue of the transition matrix P.

\paragraph{Mixing Time}

The mixing time $\tau(\varepsilon)$ is the lowest (or first) time \textit{t} at which the distribution of a chain starting from any initial state \textit{i} is below the total variation distance $\varepsilon$ of the stationary distribution $\pi$. In essence, the mixing time represents the number of transitions in the chain before the chain aligns with the long-term distribution of the overall state space (regardless of starting state). In our case, this represents the number of edits within a session before the session can be considered to reflect the overall distribution of namespace activity (regardless of the namespace in which a user starts). We calculate this metric using the formula:

\[
\tau(\varepsilon) \;=\; \min \biggl\{ t \;:\; \max_{i \in \mathcal{S}} \,\bigl\| P^t(i, \cdot) - \pi \bigr\|_{\mathrm{TV}} \leq \varepsilon \biggr\}
\]

\subsubsection{Interrupted Time-Series}

To calculate the impact that the release of ChatGPT may have had on engagement across namespaces in Wikipedia, we use two complementary Difference-in-Differences (DiD) approaches. We evaluate four metrics designed to understand impacts on engagement: edit counts, editor counts, new user counts and share of edit sessions. For edit counts, editor counts and new users, we perform a logarithmic transformation of log($Y_{it}$+1) where Y is the monthly outcome for namespace $i$ in month $t$. This transformation is intended to allow comparison between namespaces with otherwise highly diverse levels of participation. We perform no transformation for edit session shares as these are already relative percentages.

We centre time at the intervention month which we define as December 2022 following the launch of ChatGPT on the 30th of November 2022. We define $time0_t$=$t$-$t_0$ such that the intervention is a dummy variable 0 where $time0_t$ is less than 0 and 1 where it is 0 or more. Based on this, in November 2022, $time0$ is -1 and December 2022 has $time0$ of 0.

\paragraph{Level DiD-ITS}

To capture the immediate shock or change that the release of ChatGPT may have had on Wikipedia engagement, we fit the following model:

\begin{align*}
Y^{*}_{it} \;=\;& \; \alpha
+ \beta_1\,\texttt{time0}_t
+ \beta_2\,\texttt{intervention}_t
+ \beta_3\,\texttt{posttrend}_t \\
&+ \sum_{j\ne 0}\gamma_j\,\mathbf{1}[i=j]
+ \sum_{j\ne 0}\delta_j\,\mathbf{1}[i=j]\cdot \texttt{intervention}_t \\
&+ \sum_{m=2}^{12}\mu_m\,\mathbf{1}[\text{month}=m]
+ \mathbf{X}_{it}^{\top}\boldsymbol{\theta}
+ \varepsilon_{it}
\end{align*}

Where $\mathbf{1}[i=j]$ are namespace indicators with $i{=}0$ as the reference. We interpret $\beta_2$ as the immediate change in the level of engagement for namespace 0 in December 2022 with $\delta_j$ as the DiD level effect for a namespace $j$ relative to the baseline namespace 0. For those inputs with logarithmic transformations, we report effect sizes by conversion through $\%\Delta \approx 100\big(e^{\hat{\beta}}-1\big)$. For session shares, coefficients are reported as percentage points.

\paragraph{First-difference $\Delta{y}$ DiD for long-term effects}

To identify longer-term post-period effects and to allow for differences between namespaces, we estimate a DiD model on the first difference of the transformed outputs using the following model:

\[
\Delta Y^{*}_{it}
= \pi_0
+ \pi_1\,\texttt{intervention}_t
+ \sum_{j\ne 0}\kappa_j\,\mathbf{1}[i=j]
+ \sum_{j\ne 0}\lambda_j\,\mathbf{1}[i=j]\cdot \texttt{intervention}_t
+ \sum_{m=2}^{12}\rho_m\,\mathbf{1}[\text{month}=m]
+ \Delta\mathbf{X}_{it}^{\top}\boldsymbol{\phi}
+ u_{it}
\]

Where $\pi_1$ is the post gradient - pre gradient for namespace 0 and $\lambda_j$ is the additional change in the gradient observed for a namespace $j$ relative to namespace 0. Once again, we report transformed measures through conversion per month and report gradient changes in percentage points per month. 

\section{Results}

We detail each of our research questions in the order set out in the introduction. We begin by exploring how coordination has changed relative to article editing before exploring how behaviours and transitions at the edit session level reveal specialisation across Wikipedia. Finally, we present the results of our Difference-in-Differences analysis to detail how LLMs are reshaping engagement in Wikipedia's articulation spaces.  

\subsection{RQ 1 -- How has the scale of coordination changed relative to content editing over Wikipedia’s history?}

\begin{figure}[ht]
  \centering
  % First row
  \begin{subfigure}{0.48\linewidth}
    \centering
    \includegraphics[width=\linewidth]{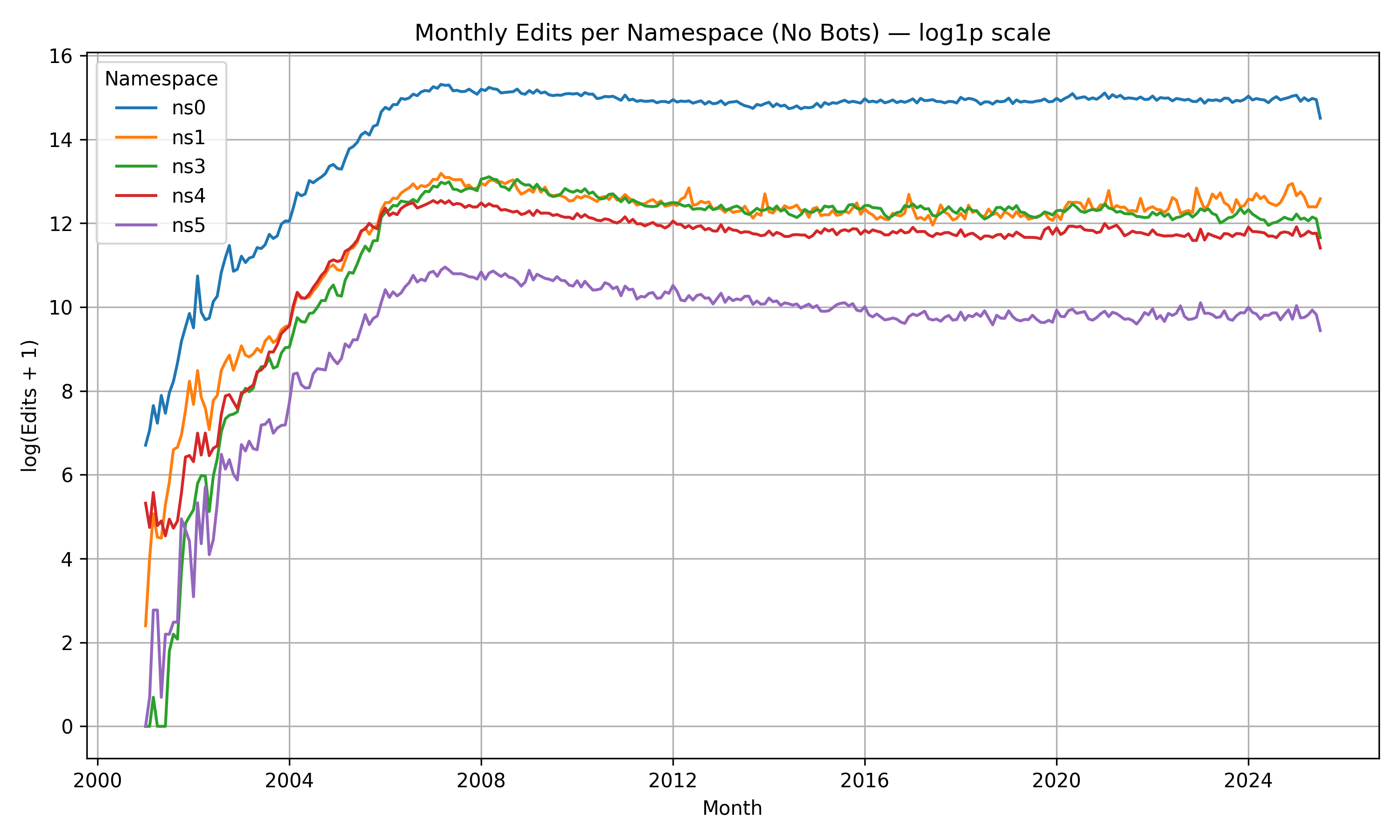}
    \caption{Monthly Edits per Namespace}
    \label{fig:monthly_edits}
  \end{subfigure}
  \begin{subfigure}{0.48\linewidth}
    \centering
    \includegraphics[width=\linewidth]{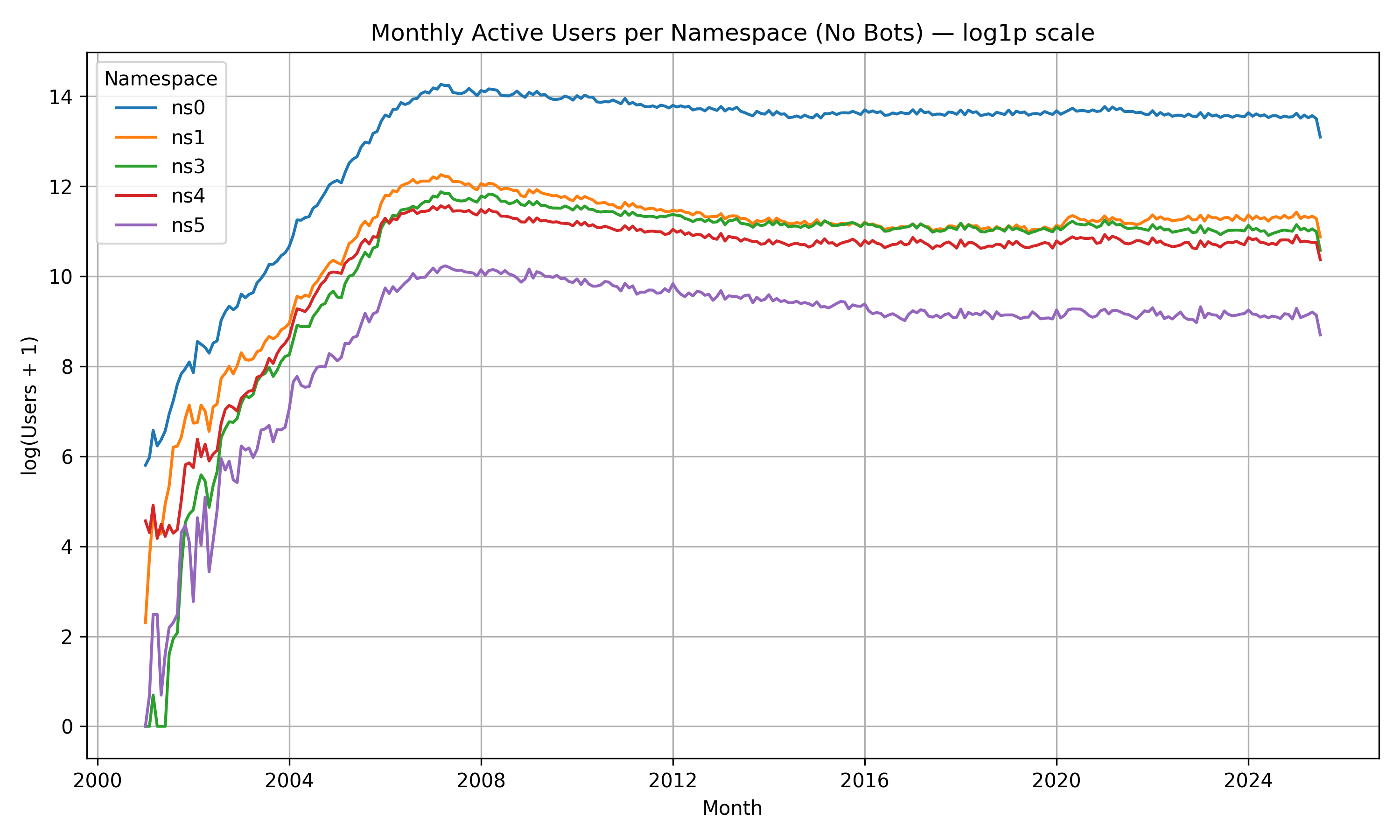}
    \caption{Monthly Users per Namespace}
    \label{fig:monthly_users}
  \end{subfigure}
  \caption{Longitudinal Analysis of Edits (Left) and Users (Right) per Namespace}
  \Description[Longitudinal analysis of Edits and Users per Namespace.]{Longitudinal analysis of Edits and Users per Namespace. Edits and Users have fallen in all namespaces at a broadly similar rate to Namespace 1.}
\end{figure}

\begin{figure}[ht]
  \centering
  % First row
  \begin{subfigure}{0.48\linewidth}
    \centering
    \includegraphics[width=\linewidth]{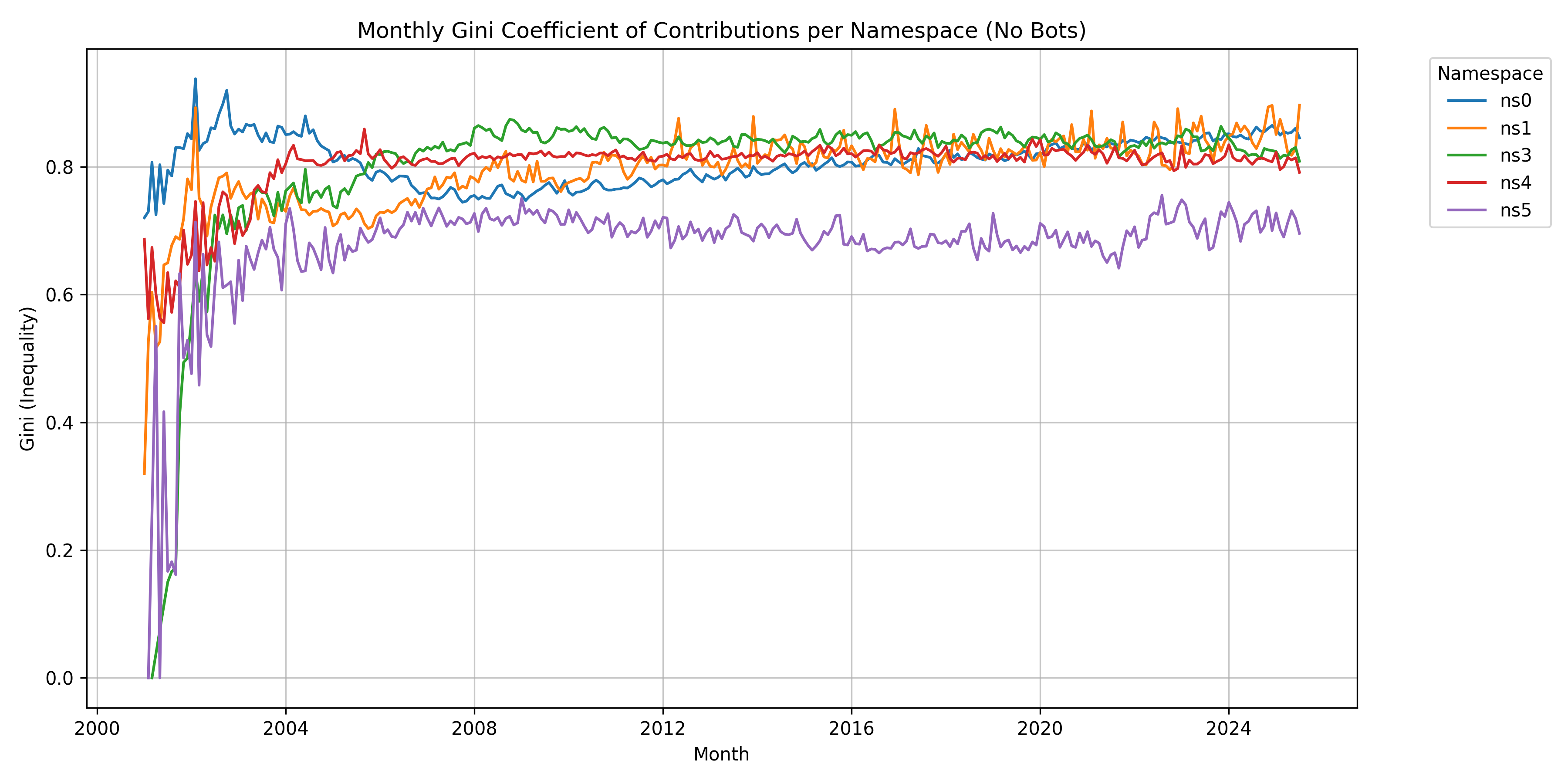}
    \caption{Gini Coefficient of Monthly Contributions}
    \label{fig:gini}
  \end{subfigure}
  \begin{subfigure}{0.48\linewidth}
    \centering
    \includegraphics[width=\linewidth]{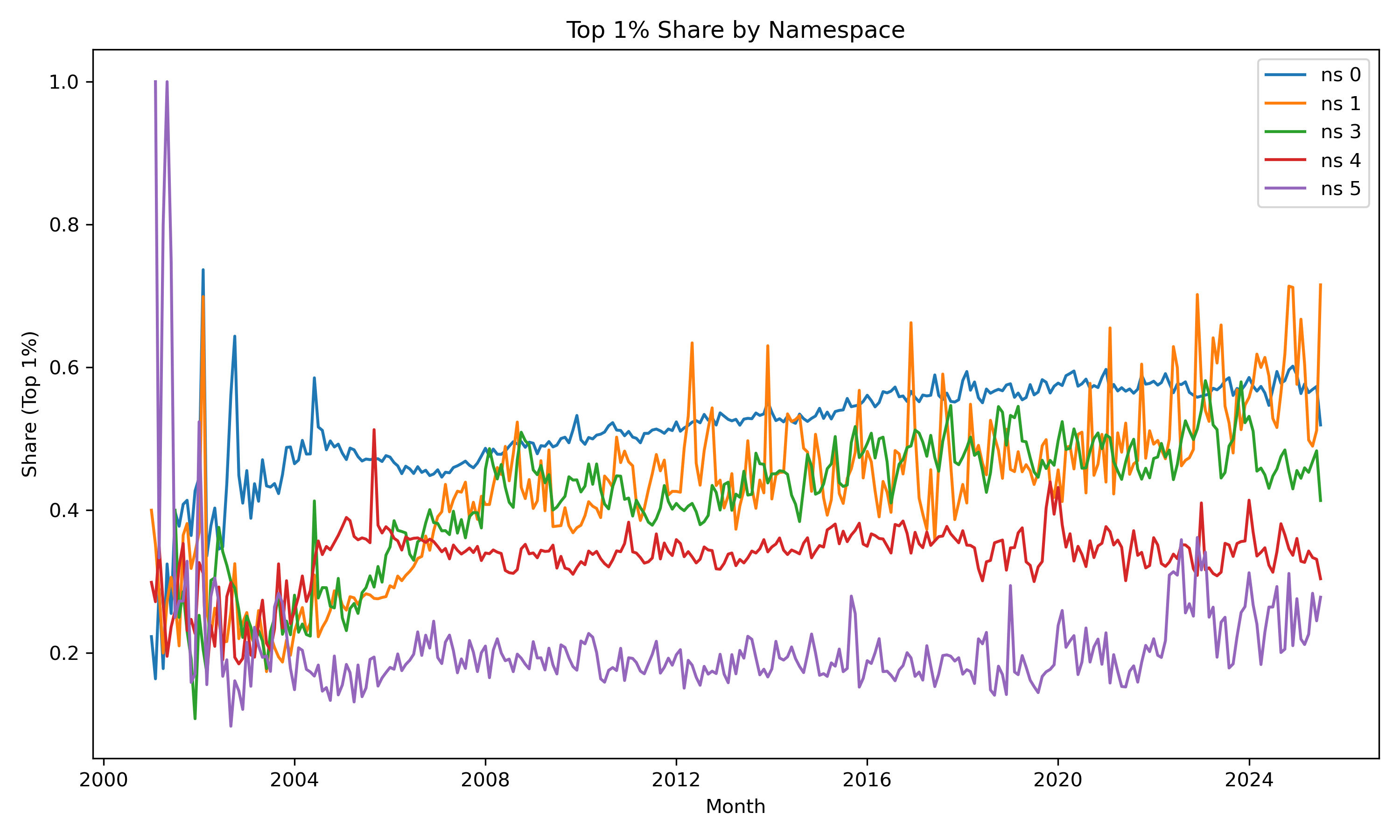}
    \caption{Share of Edits Made by Top 1\% of Wikipedia Editors}
    \label{fig:top1share}
  \end{subfigure}
  \caption{Gini Coefficient and Share of Edits Made by Most Active Wikipedia Editors Each Month as a Proportion of Total Changes.}
  \Description[Gini Coefficient and Share of Edits Made by Top 1\% and Top 10\% of Users]{Gini Coefficient and Share of Edits Made by Top 1\% of Users. Inequality has increased in all namespaces over time and the share of edits made by the top 1\% of users has similarly grown over time.}
  \label{fig:topshare}
\end{figure}

\subsubsection{Edit Counts} Monthly edit counts for all namespaces can be seen in Figure \ref{fig:monthly_edits}. Edit counts initially grew rapidly before reaching their peak around 2007. We calculated percentage changes based on the logarithm of edits (plus one) for each namespace. Article discussions experienced the lowest decline (-45.41\%) followed by main namespace articles (-55.35\%), namespace 4 (-67.81\%), 3 (-76.67\%) and the largest decline was seen in namespace 5 (-78.01\%). On the whole, coordination namespaces have experienced a larger decline than main article edits, although the similar decline in namespace 1 and 0 suggests that the level of direct coordination required to maintain Wikipedia has remained largely consistent.

%\subsubsection{Edit Counts} A chart of monthly edit counts for all namespaces can be seen in figure \ref{fig:monthly_edits}. In the earliest days of Wikipedia, edit counts rapidly grew reaching their peak around 2007 towards the end of Wikipedia's initial growth period. We calculate the percentage change based on the logarithm of edits plus one for each namespace. Namespace activity declined most quickly in namespace 5 with a  -78.01\% change from its peak, followed by namespace 3 with a -76.67\% change. A similar decline was seen in namespace 4 with a -67.81\% change. However, namespace 1 experienced the lowest decline with -45.41\%, lower even than namespace 0 with -55.35\%.

%While on the whole the coordination namespaces experienced a larger decline than the main article edits, we highlight that even namespace 1 participation has declined by almost half. Even so, the similar decline in namespace 1 and namespace 0 suggests that on the whole, the level of direct coordination required to maintain Wikipedia has remained largely consistent and perhaps even increased relative to article editing activity. Nevertheless, it is clear that other forms of coordination are experienced a sharper decline.

\subsubsection{Active Users} Similarly to edit counts, monthly active user counts have declined across namespaces. This decline was lower in namespace 0 (-68.99\%) followed by namespaces 4 (-69.90\%), 3 (-72.76\%), 1 (-74.97\%) and 5 (-78.43\%). We note a significant disparity in the rate of decline for users relative to edits in namespace 1. This would suggest that although the number of users in these namespaces has fallen, they have taken on a greater volume of work and thereby partially mitigated a fall in edit counts.

%Similarly to edit counts, monthly active user counts have declined across all namespaces. However, this decline has been similar in namespaces 1, 3  (-74.97, -72.76 percentage decline respectively) with the greatest decline seen in namespace 5 (-78.43\%). Namespace 0 has experienced the slowest decline (-68.99\%) with a similar rate of decline seen in namespace 4 (-69.90\%).

%While these rates of decline are somewhat similar in most cases, it is notable that they are larger for all namespaces other than namespace 3. In particular, we note that there is a significant disparity in the rate of decline for users relative to edits in namespace 1. This would seem to suggest that although the number of users in these namespaces is falling, they are taking on a greater volume of work and in so doing, partially mitigating any fall in edit counts. To better understand this, we explored measures of equality, particularly the Gini Coefficient.

\subsubsection{Equality} Figure \ref{fig:gini} shows the Gini Coefficient of contributions across namespaces each month, corresponding to the equality of edit counts across editors. On launch, the coefficient was low, with editing work distributed relatively evenly among editors. However, equality increased quickly reaching a peak of $\sim$0.94 (approaching perfect inequality) in 2002 for the main namespace edits. From 2005 onwards, coefficients have remained relatively stable at $\sim$0.8 for namespaces 0 through 4 and $\sim$0.7 for namespace 5. We further plotted the share of edits completed by the top 1\% of active users in each namespace each month as shown in Figure \ref{fig:topshare}. This has grown in all namespaces except for namespace 4, suggesting that the most active users have taken on increasing levels of editing and coordination work.

\begin{figure}[ht]
  \centering
  % First row
  \begin{subfigure}{0.48\linewidth}
    \centering
    \includegraphics[width=\linewidth]{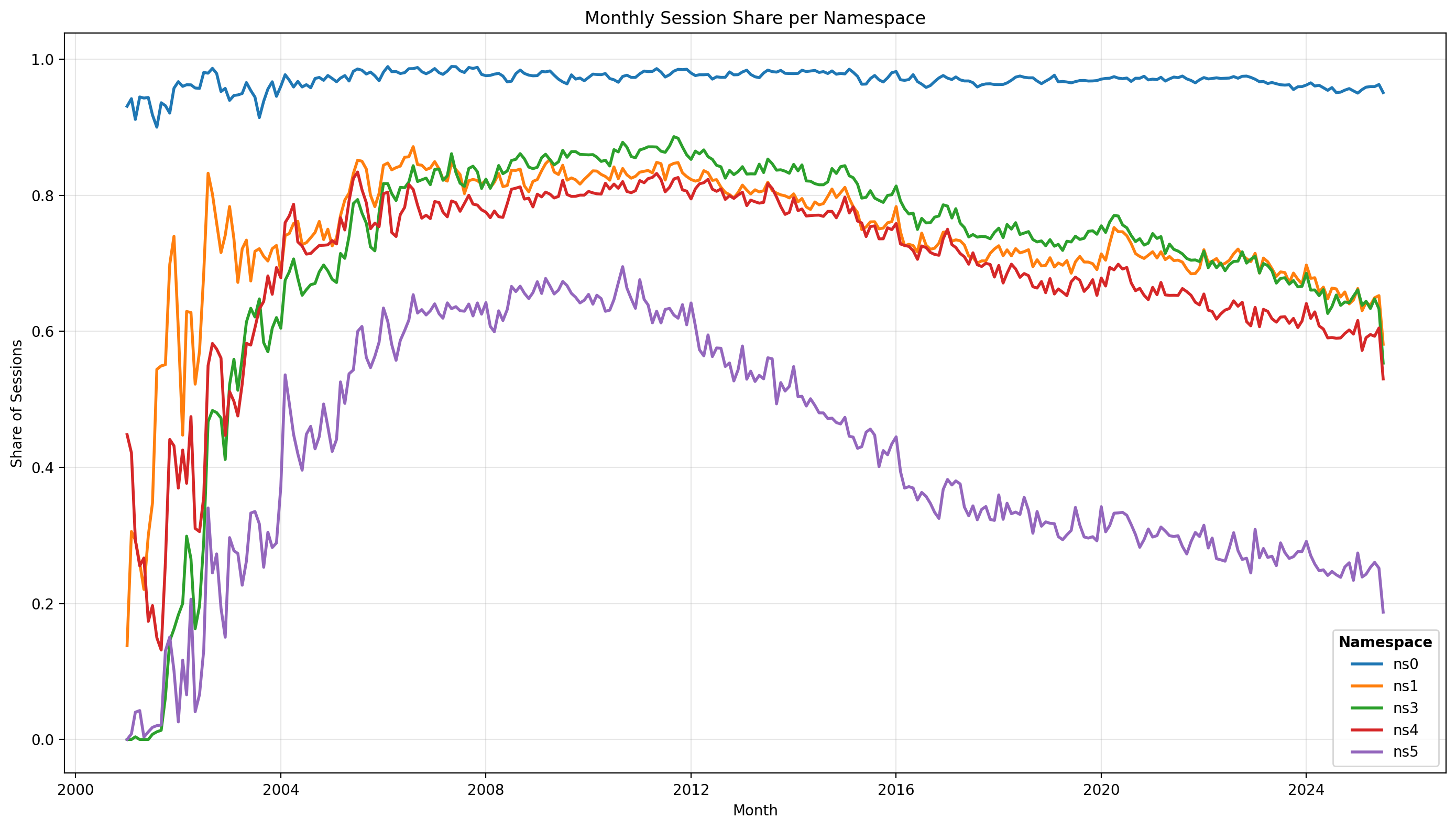}
    \caption{Monthly Share of Sessions per Namespace}
    \label{fig:monthly_share}
  \end{subfigure}
  \begin{subfigure}{0.48\linewidth}
    \centering
    \includegraphics[width=\linewidth]{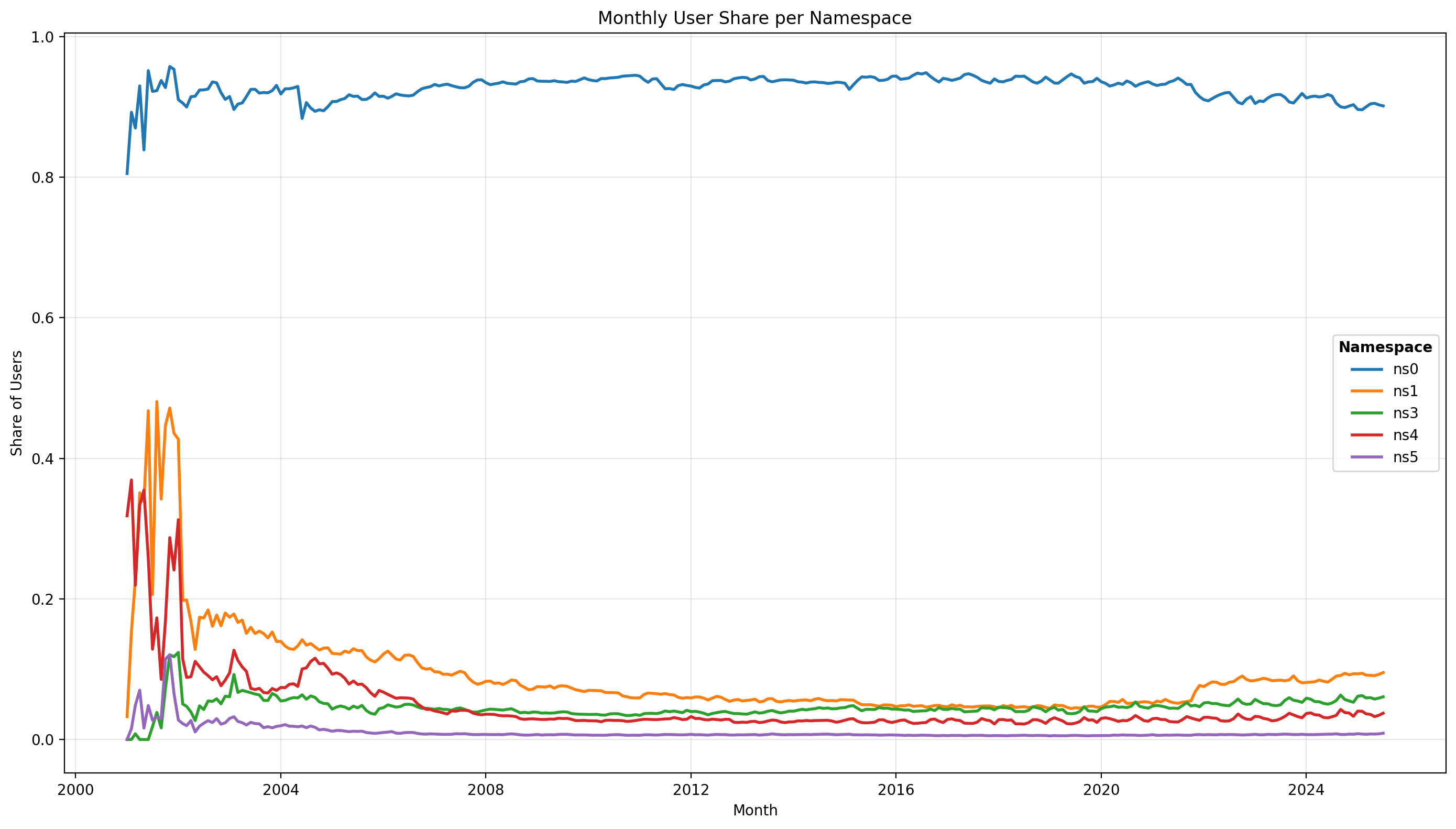}
    \caption{Monthly Share of Active Users per Namespace}
    \label{fig:month_user_share}
  \end{subfigure}
  \caption{Monthly Share of Sessions (Left) and Active Users (Right) as Proportion of Total}
  \label{fig:session_share}
  \Description[Monthly Share of Sessions and Active Users As Proportion of Total]{Monthly Share of Sessions and Active Users As Proportion of Total. Session shares have fallen for all namespaces, while active user shares have remained relatively static.}
\end{figure}

\subsubsection{Share of Sessions} Figure \ref{fig:monthly_share} details the share of sessions per month which involve edits to each namespace. Since sessions can involve multiple namespaces, shares do not sum to 1. Perhaps unsurprisingly, namespace 0 attracts the highest proportion of edit sessions and engagement in this namespace has remained consistent over time. However, only 99\% of sessions involve edits to the main namespace, suggesting the existence (albeit rare) of edit sessions focused exclusively on other namespaces.

The share of sessions involving namespaces outside of the main article space has fallen significantly over time. At their peak in 2011, the vast majority of sessions involved participation in the talk, user talk and governance namespaces (85\%, 87\% and 83\%) with 67\% of all sessions involving engagement in governance discussion. By 2025, this had fallen to 65\%, 63\% and 61\% respectively with 25\% of sessions involving engagement in governance discussion. Conversely, in namespace 0 the proportion of sessions fell from 99\% to 96\%.

%This is a much more significant decline than monthly edit counts suggest. We believe this has been balanced by the active editing community taking on a larger proportion of the work (as demonstrated by the growing Gini-coefficient). 

%As a relative decline, this is much more significant than the monthly edits suggests. We believe that this decline in session activity has been balanced by the active editing community taking on a larger proportion of the work as demonstrated by the growing Gini-coefficient. In essence, then, as the proportion of edit sessions involving contributions to these namespaces has fallen, the number of edits to these namespaces within those shrinking sessions has grown in response. This suggests that the community have taken on an increasing workload to ensure that articulation work continues.

%In contrast with the main namespace, the share of sessions with edits to all other namespaces has fallen over time, with namespace 5 (governance discussions) showing the most dramatic fall. Engagement in namespaces beyond namespace 0 as a share of the total sessions peaked in the 2010-2012 period, and while there is some degree of volatility, each namespace attracts a substantially lower proportion of edits in 2025 than it did previously. This phenomenon is not seen in the main article namespace.

\subsubsection{Share of Active Users} Figure \ref{fig:month_user_share} shows the monthly share of active users for each namespace. As with the session shares, we see evidence of a decline in the share of active users who participate in each namespace. At their peak, talk, user talk and governance namespaces attracted engagement from approximately 10\% of all active users (9.7\%, 9.3\% and 9.7\% respectively) while governance discussions attracted 7\% of active users. By June 2025, user talk attracted just 5.6\% with 3.5\% active in governance and 0.8\% active in governance discussions. Talk participation somewhat avoided this decline, falling only slightly to 9.2\% and actually increasing somewhat in the years following 2022.

\begin{figure}
    \centering
    \includegraphics[width=0.5\linewidth]{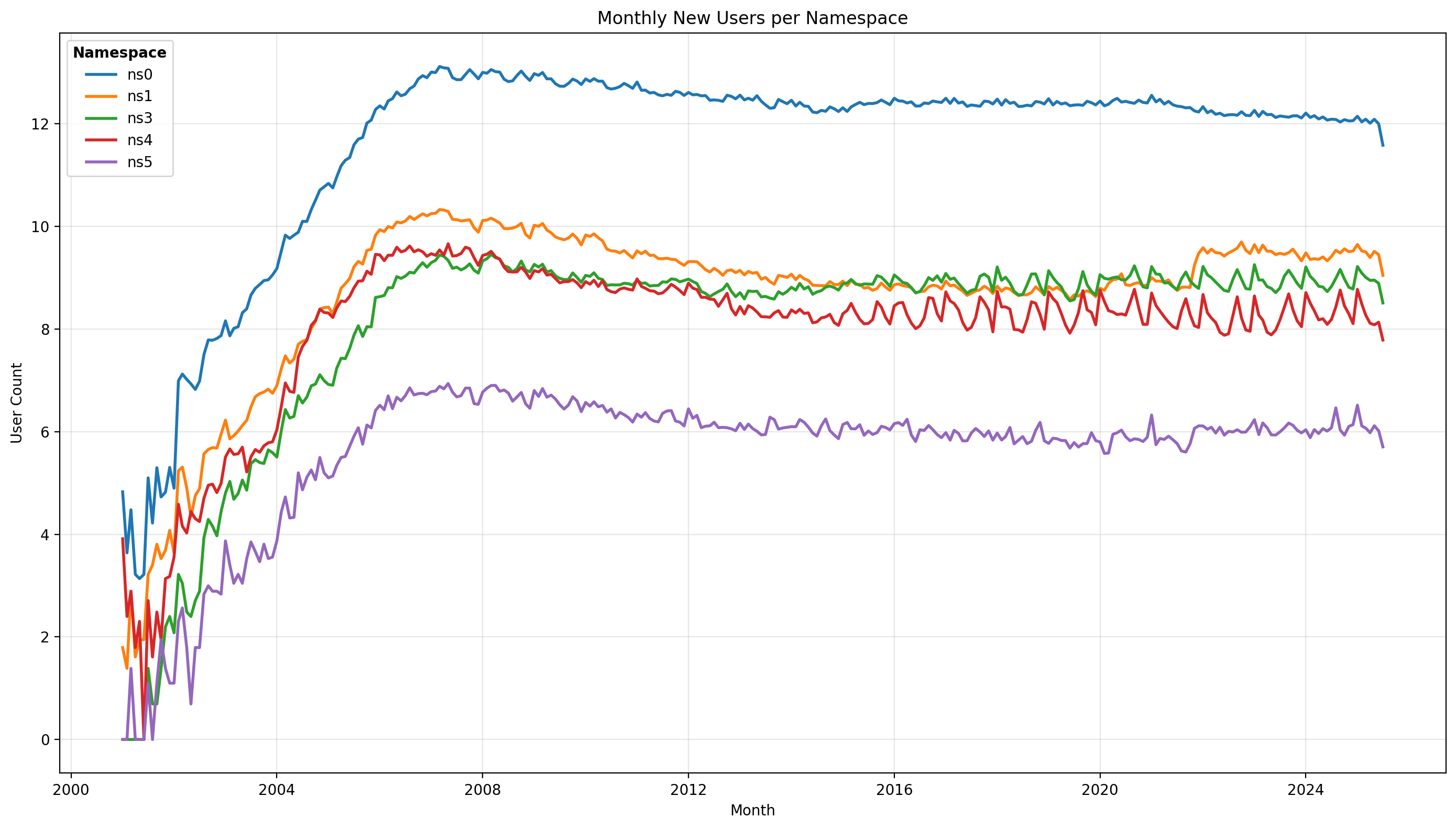}
    \caption{New Users per Namespace. We define New Users as Users Contributing to a Given Namespace in Their First Month of Wikipedia Editing}
    \label{fig:new}
    \Description[New Users per Namespace.]{New Users per Namespace. We define New Users as Users Contributing to a Given Namespace in Their First Month of Wikipedia Editing. New Users have Declined for All Namespaces.}
\end{figure}

\subsubsection{New Users} Finally, Figure \ref{fig:new} shows the number of new users contributing to each namespace over time. We define a new user as any user in their first month of Wikipedia editing regardless of namespace edited. New user counts have declined across all namespaces over time, although this has levelled out for all coordination namespaces and newcomer participation in talk has actually increased since early 2022. The rate of decline in new user participation has been greater in the main edit namespace. Nevertheless, we caution that our analysis considers only the edit paths of new users and accounts for neither the frequency nor length of time for which individuals contributed. 

%\subsubsection{Summary} As well documented Wikipedia edit counts and active users have been on the decline since 2008. Our findings demonstrate that this applies to all namespaces and that the relative decline in the user talk, governance and governance discussions has been greater than that in the main article namespace. While this decline has been partially mitigated by users taking on a greater share of the necessary work, this phenomenon has largely been restricted to a small core of highly active users. All namespaces continue to attract new users, but the share of edit sessions which involve edits to namespaces beyond the main article space has declined alongside monthly edit counts.

%On the whole, Wikipedia editing and edit counts have remained relatively consistent across namespaces. The relative popularity of each namespace has remained fixed since the peak in engagement in 2007. Moreover, engagement in each namespace has fallen at a broadly consistent rate. This suggests that the coordination overheads and articulation work conducted to support article creation have not substantially changed across Wikipedia's lifespan. However, in contrast to overall engagement, the share of activity completed by the editing community has been highly volatile and has varied across namespaces. Talk participation has become particularly specialised over time with a consistent growth in the inequality in talk participation. 

\subsection{RQ 2 -- How do session-level editing behaviours reveal specialisation and transitions across collaboration spaces?}

To understand how edit patterns and transitions between Wikipedia namespaces have evolved over time, we plot and present Markov metrics. Since different namespaces launched on different dates, we report all metrics from the start of the second year of Wikipedia (January 2002) at which point all of the analysed namespaces had been created.

%To understand how edit patterns and transitions between Wikipedia namespaces have evolved over time, we plot four Markov metrics: Stationary-Weighted Self-Transition Probability (stickiness); Spectral Gap; Mixing Time and Entropy. Not every namespace existed from the very start of Wikipedia, and an exploratory analysis of the gathered data showed that many namespaces saw little to no participation in Wikipedia's first year. This made calculating chains misleading or even impossible for some monthly periods within this year. We therefore opted to remove this year from our analysis, and we instead report all metrics from the start of Wikipedia's second year: January 2002.  A set of time-series charts showing change in metrics over time is provided in Figure \ref{fig:markov}.

\begin{figure}[ht]
  \centering
  % First row
  \begin{subfigure}{0.48\linewidth}
    \centering
    \includegraphics[width=\linewidth]{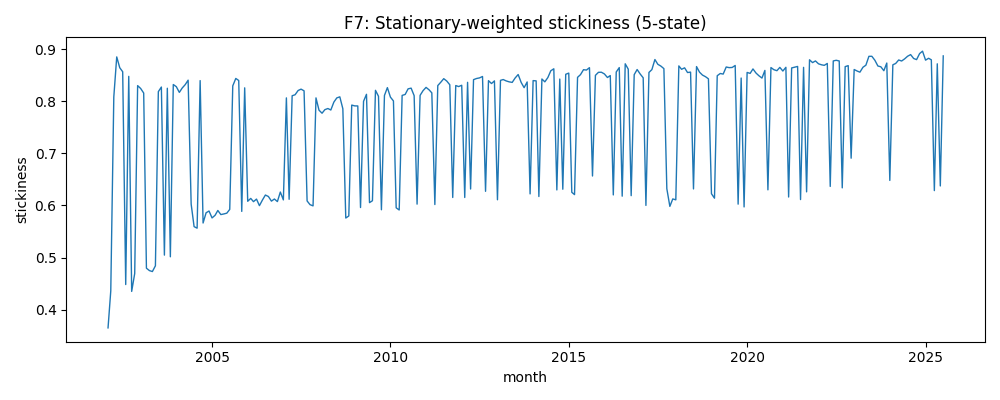}
    \caption{Stationary-Weighted Self-Transition Probability (Stickiness) Over Time Across Edit Session Chains}
    \label{fig:stickiness}
  \end{subfigure}

  \vspace{0.5em} % space between rows

  % Second row
  \begin{subfigure}{0.48\linewidth}
    \centering
    \includegraphics[width=\linewidth]{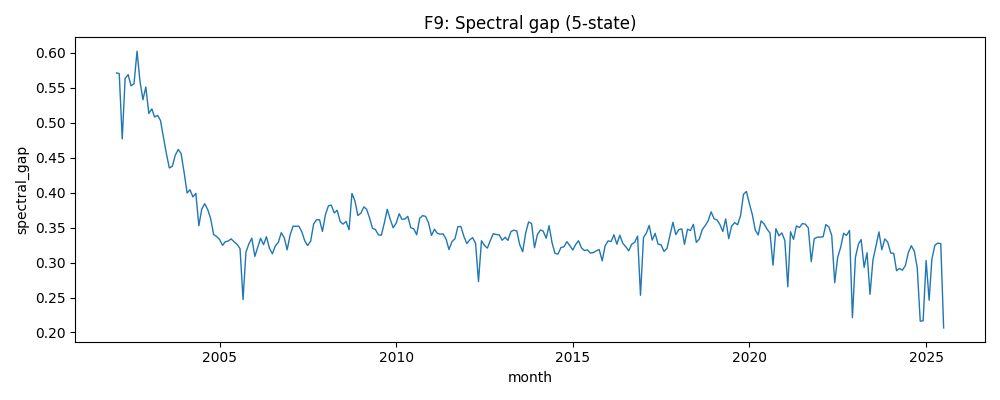}
    \caption{Spectral Gap Over Time Across Edit Session Chains}
    \label{fig:spectral}
  \end{subfigure}\hfill
  \begin{subfigure}{0.48\linewidth}
    \centering
    \includegraphics[width=\linewidth]{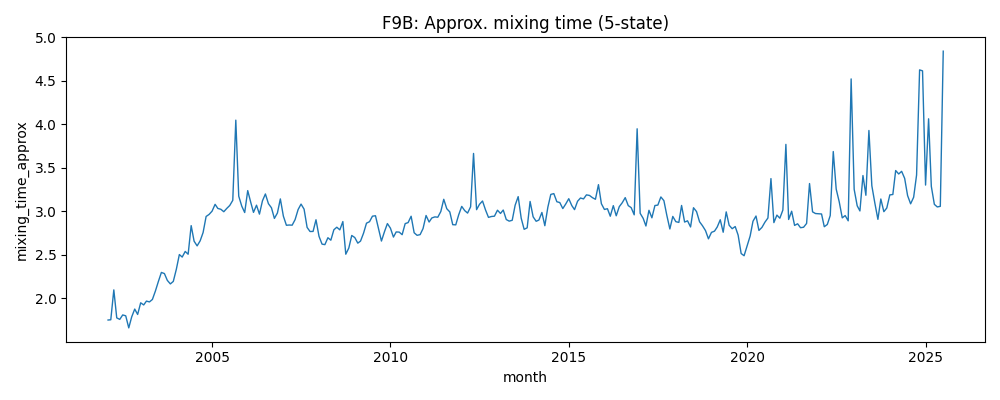}
    \caption{Mixing Time Over Time Across Edit Session Chains}
    \label{fig:mixing}
  \end{subfigure}
  \caption{Markov Metrics for Edit Sessions Over Time}
  \label{fig:markov}
  \Description[Markov Metrics for Edit Sessions Over Time]{Markov Metrics for Edit Sessions Over Time including Stationary Weighted Self-Transition Probability (Stickiness); Entropy Rate; Spectral Gap and Mixing Time.}
\end{figure}

\subsubsection{Self-Transition Probability} The `stickiness' over time can be seen in Figure \ref{fig:stickiness}. This metric has peaked in recent years reaching a high of 0.89 in November 2024. However, it is highly volatile and the probability has repeatedly fallen to approximately 0.6 throughout the lifespan of Wikipedia. Further analysis suggests that while this fall happens at least once a year, there is no consistency in the period between -- or the months which -- the observed fall. Periods where a fall is observed are typically fleeting and consist of just a single month, although a four month fall was observed between October 2017 and January 2018. 

%Following highly generalist editing behaviours in the first year of Wikipedia's life and despite some brief periods where stickiness has fallen, the majority of chains involve persistent edits within a specific namespace. The extent of this stickiness has gradually increased since 2009 having reached its current height in 2024, but is likely to trend higher if this observed phenomenon persists. This likely reflects a majority of users focusing exclusively on a single -- or small number of -- namespace(s), particularly the main article space which consistently attracts a high share of sessions as demonstrated previously. 

\subsubsection{Spectral Gap} Figure \ref{fig:spectral} shows the spectral gap across all edit chains over time. Between 2006 and approximately 2019, the spectral gap remained relatively stable despite some volatility before declining sharply following a peak in 2020 aligning with the start of the global coronavirus pandemic. It has continued to fall since and reached its lowest ever point in July 2025.

\subsubsection{Mixing Time} Figure \ref{fig:mixing} shows that the complementary mixing time has trended correspondingly higher, rising from under 2 months in the earliest days of Wikipedia to over 5 months in 2025. Taken together, these metrics suggest that edit sessions have become increasingly specialised following Wikipedia's initial period of growth, reaching a broad equilibrium that held until 2020. This trend briefly reversed at the time of the coronavirus pandemic where sessions became more generalist, perhaps reflecting increased coordination demands or more free time for editors, but specialisation has since resumed and intensified.

\subsubsection{Implications} Our findings suggest that editing behaviours have become increasingly compartmentalised over time. Once editors enter a namespace, they are increasingly likely to remain there and when they do transition, they tend to do so in more predictable ways. Taken alongside our analysis of contribution equality, this suggests an editing ecosystem that is increasingly siloed and dominated by a small minority of editors.

\subsection{RQ 3 -- How has the emergence of LLMs reshaped engagement in articulation spaces compared to content spaces?}

For research question three, we explored the impact that the availability and potential use of LLMs may have had on engagement across namespaces. Using a DiD analysis, we compare  each namespace against namespace 0 (main content) as a baseline control and report all coefficients and statistical significance ratings accordingly. 

\subsubsection{Edit Counts} The release of ChatGPT was associated with an immediate reduction in monthly edit counts relative to namespace 0 in namespaces 4 (p\textless0.001; effect =-22\%) and 5 (p\textless0.001; effect =-25\%) with a very weakly statistically significant increase relative to the baseline in namespace 1 (p=0.033; effect =+10.8\%). We find no evidence of a long-term trend in any of the namespaces. Model fit was relatively high for the Difference-in-Differences model ($R^2$=0.827) but much lower in the slope comparison ($R^2$=0.081). This is typical for first difference models, but we can confidently conclude that we find no evidence of long-term changes.

\subsubsection{Editors} Similar to edit counts, the release of ChatGPT was associated with an immediate reduction in monthly editor numbers relative to the baseline of namespace 0 in namespaces 4 (p=0.006; effect size=-12.9\%) and 5 (p\textless0.001; effect size = -21.1\%). Additionally, a very weakly statistically significant reduction was seen in namespace 1 (p=0.037; effect size = -8.0\%). When observing longer-term trends, however, we find no evidence of a statistically significant trend in any namespace. Model fit was relatively high for the Difference-in-Differences model ($R^2$=0.82), while it was significantly lower for the slope comparison ($R^2$=0.11).

%As with edit counts, the release of ChatGPT was associated with an immediate reduction in editor numbers in namespaces 3 (\textless0.001; -0.7108), 4 (0.007; -0.216) and 5 (0.001; -0.377). No statistically significant trend was identified for the period following the release of ChatGPT suggesting that there was no long-term impact on namespace editing beyond what might be expected based on fluctuations in namespace 0. Model fit was again high ($R^2$=0.826).

\subsubsection{New Editors} New user counts grew significantly relative to namespace 0 in namespace 1 (p\textless0.001; effect size=+43.5\%) and 3 (p\textless0.001; effect size=+52.9\%). Conversely, a significant fall was seen in namespace 4 (p=0.007; -19.6\%) and no statistically significant adjustment was observed in namespace 5. Once again, no long-term trend was observed beyond that of the baseline. Model fit was of a similar strength ($R^2$=0.869; 0.113)

\subsubsection{Share of Sessions} We find evidence of a strong decline in the share of edit sessions which involved changes to each namespace. All namespaces saw a statistically significant reduction in edit share on the release of ChatGPT relative to namespace 0 (p\textless0.001; effect-size \textit{1}: -7.21\%, \textit{3}: -5.32\%, \textit{4}: -8.70\% \& \textit{5}: -16.62\%). Only namespace 3 showed a statistically significant long-term change and this was weak (p=0.041; effect-size=-0.63\% per month). Model fit was weaker in comparison with other models ($R^2$=0.661; 0.023). 

%\subsubsection{Summary} Our findings suggest that the release of ChatGPT had an immediate impact on coordination work within Wikipedia. The number of editors and volume of edits completed in the governance and governance discussion namespaces fell (although edit activity increased in namespace 1). Conversely, a higher number of new users were attracted to talk and user talk namespaces (unlike governance participation). We found no limited evidence of any long-term effects with the exception of a small long-term reduction in the share of edit sessions involving edits to user talk.

\section{Discussion}

\subsection{Longitudinal Trends}

Our analysis shows that the prevalence of talk and user talk coordination relative to article edits has remained broadly stable. Conversely, governance engagement has diminished (particularly discussion) and overall engagement in all namespace activities is on the decline. While this has been partially mitigated by an increased workload among the top 1\% of editors, we also see a shift towards specialisation in collaborative activities across Wikipedia. Core contributors within namespaces are therefore increasingly assuming responsibility for governance, article maintenance or discussion co-ordination. 

Prior research has highlighted the tendency of new users to spread their contributions across many namespaces and thereby potentially hasten their departure \cite{qin2014latent}. At the same time, contributions have become more centralised \cite{halfaker2013rise, teblunthuis2018revisiting} due to declining engagement as well as low newcomer retention. To some extent, then, this increased specialisation and reduction in effort across namespaces is to be expected.

Nevertheless, these trends matter for Wikipedia because they increase the potential vulnerability of its collaboration system. If core contributors choose to leave the platform, a large portion of work could be disrupted, placing additional pressure and workload to remaining users. This may lead to saturation of work in the community, taking attention away from certain coordination activities, reducing cross-boundary collaboration as contributors limit their attention to specific namespaces or areas of work. Prior evaluation of Indian-language Wikipedia communities has highlighted the challenges for engagement that can be caused by the reliance on a small community of users for governance \cite{khatri2022social}. The loss of editors not only reflects a loss of volunteer manpower, but also a loss of critical expertise and perspectives \cite{das2021expertise}. Even should editors choose not to leave, they may risk burn-out from the increasing workload, while the shrinking community of \textit{experienced} editors in a namespace may lead to bottlenecks forming in decision making or oversight tasks. 

This observation of a small core of users who take increasing responsibility for managing an ever growing coordination workload is not unique to Wikipedia. For example, moderators in the online community Reddit take on a high workload of predominantly invisible work with high levels of heterogeneity \cite{li2022all}. A similar phenomenon has been observed in open source software communities where most coordination tasks are experienced by a small core of participants \cite{joblin2017evolutionary}. However, Wikipedia notably differs in that many user \textit{could} engage in these processes -- unlike in Reddit, there are no power dynamic restrictions that limit the pool who can coordinate and all tasks are visible to all users. Even so, we would expect these observations and implications to be applicable to other large scale peer-production communities beyond the Wikipedia context.

\subsection{Namespace Transitions}

Our findings suggest that engagement within and across namespaces has remained reasonably consistent for much of Wikipedia's lifespan. While engagement has consistently fallen in all namespaces with the exception of namespace 3, this disparity was low, suggesting that the volume of articulation work and the coordination overheads associated with editing have largely remained consistent. Nevertheless, our findings suggest that on the whole, editors have become increasingly siloed within specific namespaces as Wikipedia has evolved. 

One potential concern this surfaces is that decisions and suggestions from one namespace may fail to transition to other namespaces as the community of users who could act on them shrinks. For example, as previously noted, article talk pages are regularly used in English Wikipedia to gather and discuss suggested revisions to articles \cite{bipat2018we}, but as \citet{im2018deliberation} have highlighted,  the effectiveness of such discussions is limited by the failure of editors to close (and thereby act) on discussions. Choosing not to edit a namespace does not, of course, indicate that a user is not aware of it, but prior research suggests most users only choose to read the main namespace \cite{ren2023did}. In essence, there is a risk that as the community becomes more specialised and siloed, it also becomes fragmented. This could negatively impact both Wikipedia's growth and the quality of articles and further exacerbate the limited change in quality across article lifespans \cite{das2022quality}.

\subsection{Impact of LLMs}

While we found evidence of an immediate `shock' upon the release of ChatGPT associated with a  fall in engagement across namespaces, we found no evidence of long-term changes in engagement trends for articulation namespaces (beyond what might be expected based on changes in main article edits). Being purely quantitative and descriptive, our analysis does not specifically allow us to identify why such a shock might have been experienced. However, we note relevant observations from prior work focusing on the impact of LLMs on the main article namespace.  \citet{wagner2025death} found a significant relationship between Wikipedia readership and contributor levels, which may suggest that a fall in traffic coinciding with the release of ChatGPT could have diverted effort and attention away from Wikipedia. While \citet{reeves2024exploring} found no impact on page views at an aggregate level, a more nuanced evaluation by \citet{lyu2025wikipedia} suggests greater impacts in newer pages than others as well as topic-related facts. Our analysis was by necessity topic-independent, largely due to the lack of a universal topic model covering different namespaces.

On the one hand, this can be seen as a positive for Wikipedia. While our use of namespace 0 as a control means we cannot comment specifically on any impact on article edits, our findings suggest that in the longer-term, editors have continued to contribute to articulation work in discussion and governance spaces at a broadly similar rate. Nevertheless, we caution that this interpretation assumes that the level of discussion and governance remained static following the launch of LLMs. There is ample scope for LLMs to have a negative impact on the quality of Wikipedia content through, for example, introducing false or unsourced claims into article edits \cite{vetter2025endangered}.

Moreover, we highlight that impacts on individual namespaces appear to have been mixed. While namespace 1 experienced a relative increase in editing activity despite its fall in editor numbers, namespaces 4 and 5 experienced a much greater reduction in edits than their editor numbers. This is perhaps particularly surprising as in the wake of ChatGPT's release, we might expect a greater than usual focus on governance and defining policy. If the distribution of work beyond the main namespace has remained largely the same, then this might suggest that engagement in crucial correction or governance activities related to LLM use and content has either been overlooked or has drawn attention away from other namespace responsibilities.

\subsection{Future Work}

Our analysis opens several directions for future research. Firstly, since talk page use is highly language-dependent (see for example, \cite{bipat2018we}), future work could extend this work to other Wikipedia language editions. Similarly, our analysis focused only on quantiative measures and this could be extended through analysis of namespace \textit{content} to further explore usage patterns.

Thirdly, future research could use interviews to explore user motivations in moving between namespaces. Finally, while we focused on what we viewed to be the most significant and relevant namespaces, future work could benefit from exploring additional namespaces and how they contribute to Wikipedia's system of collaboration. 

%Our analysis opens several directions for future research. Firstly, extending this work to other language editions of Wikipedia would provide valuable insights into coordination practices, particularly in talk pages as their use is highly language-dependent \cite{bipat2018we}. Secondly, to complement the quantitative measures used to measure namespace usage, future work could focus on the analysis of namespace content itself to capture the subtle changes in how certain namespaces are used.

%Third, future research could further investigate the transitions between namespaces, using interviews with users to reveal ideas about their motivations to contribute in a given namespace. Exploring how and why users shift their focus of work could help gain deeper insight into cross-boundary collaboration. Finally, this study concentrated on a subset of namespaces which are most central to coordination, but further work would benefit from exploring additional namespaces and how they contribute to Wikipedia's system of collaboration. This could provide a broader understanding of how coordination and governance is organised across the platform's infrastructure.

\subsection{Limitations}

%We record edit sessions based on when they start -- some edits sit over the month boundary

%May not have captured all bots

We note four main limitations of our study. Firstly, while we have aimed to minimise the risk of bot contributions in our dataset, we recognise the risk that some bots may not be flagged or identified. Secondly, although only representing 1.28\% of edits, the decision to exclude anonymous users may have influenced engagement dynamics. Thirdly, our operationalisation of edit sessions as Markov chains is subject to the Markov property, which assumes that the future state depends solely on the present state rather than the full history of past states. We believe that a first-order Markov chain is sufficient to capture general transitions between namespaces, but note that higher-order chains may be more appropriate for analysing longer-range edit dynamics. Finally, although our definition of an edit session as one hour aligns with prior work (e.g., \cite{geiger2013using}), this may distort edits from lower intensity editors and future work could consider alternative session cut-off times.

\section{Conclusion}

In this paper, we provided a longitudinal analysis of almost 25 years of coordination in English Wikipedia. Our findings offer a nuanced perspective on the balance between coordination and production work: on the one hand, participation in most namespaces has fallen relative to article edits with participation becoming more siloed and fragmented. On the other, we observe that a dedicated core of users has mitigated the impact of these changes. Moreover, despite a decreasing focus on coordination activities, we find that LLMs have had no significant long-term effects on the balance and trajectories of coordination and production. Nevertheless, we caution that our analysis was inherently quantitative, and we do not attempt to point to any specific root causes. Moreover, we did not consider impacts on accuracy, and we do not claim that all activity in the studied namespaces represents coordination or articulation work. Even so, we believe that our findings suggest a stable balance between coordination and production, albeit one that may be somewhat fragile in the longer-term. %We also believe there is ample opportunity for future research to further explore how often-invisible articulation work supports and influences overall community productivity. 

%%\begin{acks}
%%TBD
%%\end{acks}

\bibliographystyle{ACM-Reference-Format}
\bibliography{WikiChatGPT}

%%\appendix

\end{document}